\documentclass[%
 aip,
 jcm,%
 amsmath,amssymb,
preprint,%
]{revtex4-2}

\usepackage{graphicx}
\usepackage{dcolumn}
\usepackage{bm}

\begin{document}

\preprint{AIP/123-QED}

\title{Manning-type potential induced by kink scatterings with phonons in molecular chains with hyperbolic double-well substrates}

\author{Alain M. Dikand\'e}

\email{dikande.alain@ubuea.cm.}
\affiliation{Laboratory of Research on Advanced Materials and Nonlinear Science (LaRAMaNS), Department of Physics, Faculty of Science, University of Buea PO Box 63 Buea, Cameroon.
}%

\date{\today}

\begin{abstract}
A rescaled Manning potential is obtained in the analysis of scatterings of small-amplitude excitations with a kink defect. The generic model is a nonlinear Klein-Gordon Hamiltonian describing a one-dimensional chain of identical molecules, subjected to an hyperbolic single-particle substrate potential. To account for isotope effects that are likely to affect characteristic equilibrium parameters of the molecular chain, including the lattice spacing (i.e. the characteristic intermolecuar distance) and/or the barrier height, the hyperbolic substrate potential is endowed with a real parameter whose variation makes it suitable for the description of molecular excitations in a broad range of systems with inversion symmetry. These include hydrogen-bonded molecular crystals, $\alpha$-helix proteins, long polymer chains and two-state quantum-tunneling systems in general. Double-well models with deformable profiles are relevant in physical contexts where the equilibrium configurations are sensitive to atomic or molecular substitutions, dilution, solvation and so on.
\end{abstract}

\keywords{Hyperbolic double-well potentials, kink defects, phonons, kink-phonon scatterings, Manning potential}
\maketitle

\section{\label{sec:level1}Introduction}
Double-well models have been instrumental in the investigations of dynamical properties of chemical systems exhibiting bistable configurational energies, due to a structure marked by an inversion symmetry \cite{a1,a2,a3}. Such chemical systems are very common in nature both in the living and non-living worlds, the most explored physical context being the one of chemical structures stabilized by hydrogen bonds \cite{a1,a3a,a11,a4,a5}. Hydrogen bonds are specific among intermolecular interactions in that they are chemical bonds between van der Waals
interactions that are moderate, and covalent bonds that are relatively stronger. Owing to this specific feature hydrogen bonds favor clustering of molecular units, promoting short to long-chain molecular structures such as dimers, trimers and so on. \\
Hydrogen bonds have extensively been investigated in the contexts of one-dimensional molecular crystals with chemical structures $X-H\cdots X$ among which water ($HO\cdots H-OH$), in these chemical systems proton defects resulting from ionization of the molecular chain are transferred via collective migrations across hydrogen bridges \cite{a3a}. However hydrogen bonds are not only about water, indeed they also play important role in many other systems such as the so-called hydrogen-bonded (or quantum) ferroelectrics \cite{df1,df2}, macromolecular and long protein chains. In proteins, intermolecular hydrogen bonds between complementary nitrogen base pairs (cytosine-guanine and adenine-thymine) connect two strands of DNA together \cite{a11,p1,p2,p3,p4}, conferring them a double-helical structure fundamental in the replication of the genetic code. Intermolecular hydrogen bonds are also known to affect the structure of proteins, in particular they maintain the cellulose or polymeric structure of protein chains \cite{p5,p6}.\\
Theoretical efforts have been undertaken to elucidate the mechanism of formation of charge defects as well as their collective migrations \cite{a3a,a6,a7,a8,a9} along molecular chains stabilized by hydrogen bond. In these studies, the energy landscape created by the hydrogen bond exhibits a bistable symmetry represented either by the $\phi^4$ potential \cite{a6,a10,a10a} or the double-Morse potential \cite{a3a}. While the former has a rigid profile given its two degenerate minima and barrier height that are fixed, the presence of an extra real parameter in the double-Morse potential \cite{a3a} favors a shape deformability such that positions of the two degenerate minima, and eventually the height of the energy barrier, can be adjusted to account for isotope effects. However, although the double-Morse potential is closer to the physical reality compared with the $\phi^4$ potential, it is an artificial bistable potential resulting from two Morse potentials \cite{morse} placed back to back to form a double-well function \cite{df1}. \\
Several far more realistic double-well potentials have been proposed whose shape profiles are purposefully adjustable, depending on the specific physical context at hand \cite{m1,m2}. One of them is the following family of hyperbolic potentials, whose double-well shape can be tuned very simply in distinct ways:
\begin{equation}
V_{\mu}(x)=a_{\mu}\bigg\lbrack\frac{1}{\mu^2}\sinh^2\left(\alpha_{\mu}\, x\right) - 1 \bigg\rbrack^2, \label{eq1}
 \end{equation}
where $a_{\mu}$ and $\alpha_{\mu}$ are two real quantities that depend on a single parameter $\mu$, assumed to be real and finite. Depending on specific physical contexts $a_{\mu}$ and $\alpha_{\mu}$ will be defined so that $V_{\mu}(x)$ mimics a double-well potential with either a fixed barrier height but tunable minima positions, fixed minima positions but a tunable barrier height, or simultaneously tunable minima positions and barrier height. The three different physical configurations corresponding to three different combinations of the real parameters $a_{\mu}$ and $\alpha_{\mu}$, describe three distinct physical contexts that can arise in hydrogen-bonded molecular chains and biomolecular chains in general, as a result of isotope effects.\\
In this work we are interested in the dynamics of kink defects in a one-dimensional molecular chain, assuming that the configurational energy possesses a bistable symmetry and shape deformability that can be described by the family of hyperbolic double-well potentials $V_{\mu}(x)$. Our primary objective is to construct the eigenvalue problem associated with the scatterings of kinks with small-amplitude modes forming in the kink background. In effect the kink-phonon scattering problem is relevant in the analysis of kink stability in real molecular crystals, where phonons are also present \cite{ph,ph1} besides kink defects. It will turn out that the eigenvalue problem associated with kink-phonon scattering for the hyperbolic double-well model eq. (\ref{eq1}), puts into play a scattering potential which is a rescaled form of the potential proposed by Manning in his study of the energy levels of $NH_3$ and $ND_3$ molecules \cite{man}.\\
In sec. \ref{sec:level2} the model is presented, it consists of a one-dimensional chain of linearly interacting identical molecules sitting on a background substrate. Individual molecules along the chain are bound to the substrate by a substrate potential of a double-well shape, reminiscent of an inversion symmetry characteristic of the molecular chain. Since the total Hamiltonian of the system is discrete the resulting equation of motion is analytically untractable, prompting the use of the continuum-limit approximation which yields an analytical solution featuring a kink soliton. In sec. \ref{sec:level3} the eigenvalue problem for the kink scattering with phonons is discussed. For this analysis we assume that in addition to the kink deformation,  molecular displacements also induce small-amplitude excitations in the kink background whose amplitudes are determined by a linear Schr\"odinger equation with a scattering potential created by the kink.  The study ends with sec. \ref{sec:level4} where a brief summary and concluding remarks are given.

\section{\label{sec:level2}The model and kink soliton solution}
\begin{figure}\centering
		\begin{minipage}[c]{0.34\textwidth}
\includegraphics[width=2.in, height= 1.8in]{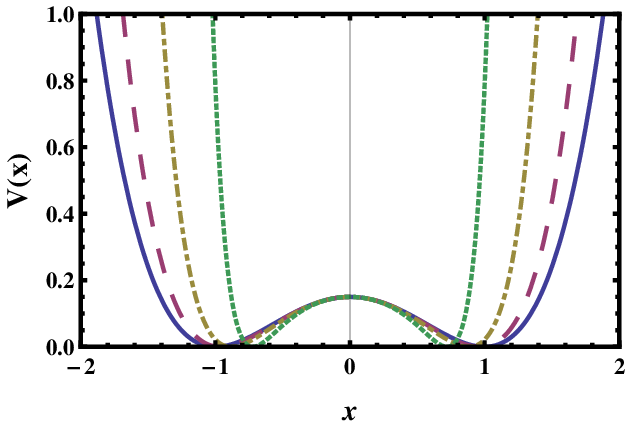}
		\end{minipage}%
		\begin{minipage}[c]{0.34\textwidth}
\includegraphics[width=2.in, height= 1.8in]{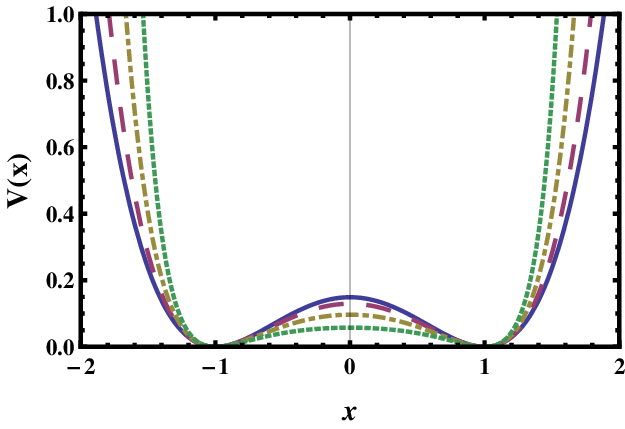}
		\end{minipage}%
		\begin{minipage}[c]{0.34\textwidth}
\includegraphics[width=2.in, height= 1.8in]{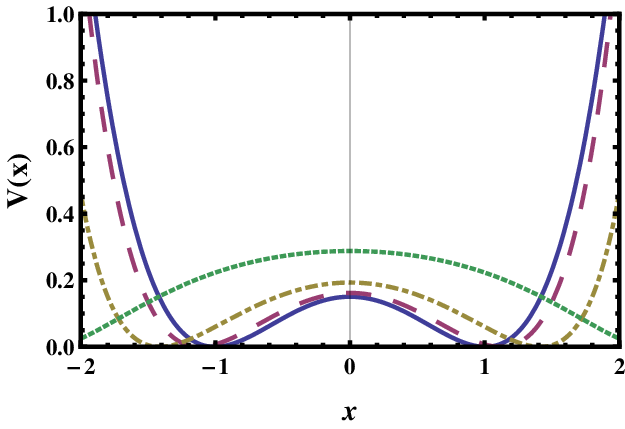}
		\end{minipage}
		\caption{Plot of the hyperbolic double-well potential $V{\mu}(x)$ for $\mu=0.1$ (solid line),
   $\mu=0.5$ (dash line), $\mu=1.0$ (dash-dotted line) and $\mu=2.0$ (dot line). From left to right graphs: tunable minima but fixed barrier, tunable barrier height but fixed minima, tunable minima and tunable barrier height.}
		\label{fig1}
	\end{figure}

Consider a molecular crystal composed of identical molecules forming a one-dimensional lattice. Molecules are assumed to interact via Hookes couplings, while individual molecules are bound to a common substrate via a one-body force provided by a double-well substrate potential. Denoting by $m$ the mass of molecules and by $\lambda$ the coupling strength between neighbor molecules, the Hamiltonian corresponding to a chain of $N$ molecular sites can be written:
\begin{equation}
H=\sum_{n=1}^{N}{\Bigg\lbrack\frac{m}{2}\biggl(\frac{\partial y_n}{\partial t}\biggr)^2 + \lambda\biggl(y_{n+1} - y_n\biggr)^2 + V_{\mu}(y_n)\Bigg\rbrack}, \label{eq2}
\end{equation}
where $y_n$ is the displacement of the $n^{th}$ molecule from its equilibrium and $t$ is time variable. The discrete Hamiltonian (\ref{eq2}) is not tractable analytically, due to discreteness that makes it difficult to apply existing mathematical methods for solving nonlinear Klein-Gordon equations \cite{dik1,dik2}. Nonetheless analytical solutions can be obtained under reasonably acceptable approximations, such as the continuum-limit approximation \cite{f4}. In this approximation it is assumed that the wavelength of molecular excitations is very large compared to the lattice spacing $d$, so large that the molecular chain behaves like a continuous medium. Therefore in this regime the discrete variable $n$ can be replaced by a continuous position $x=nd$, so we can write $y_n(t)=y(x,t)$. With these considerations and assuming the dispersion to be weak, the Hamiltonian eq. (\ref{eq2}) is transformed to the following continuous nonlinear Hamiltonian:
\begin{equation}
 H=\frac{m}{2d}\int_0^{\infty}{dx\Bigg\lbrack\biggl(\frac{\partial y}{\partial t}\biggr)^2 + c_0 ^2\,\biggl(\frac{\partial y}{\partial x}\biggr)^2 + 2\omega_0^2\,\tilde{V}_{\mu}(y)\Bigg\rbrack}, \label{hcont}
\end{equation}
where the characteristic speed $c_0$ and frequency $\omega_0$ of long-wavelength acoustic vibrations of the one-dimensional molecular chain are defined as:
\begin{equation}
c_0^2=\lambda\,d^2/m, \qquad \omega^2=a_0/m,\qquad \tilde{V}_{\mu}(y)=V_{\mu}(y)/a_0.
\end{equation}
The continuous Hamiltonian eq. (\ref{hcont}) leads to the following nonlinear Klein-Gordon equation:
\begin{equation}
\frac{\partial y^2}{\partial t^2} - c_0^2\,\frac{\partial y^2}{\partial x^2} + \omega_0^2\,\frac{\partial}{\partial y}\tilde{V}_{\mu}(y)=0. \label{eq8}
\end{equation}
In this nonlinear partial differential equation, the double-well potential $V_{\mu}(y)$ whose general expression is given by (\ref{eq1}), is defined such that:
\begin{enumerate}
 \item for:
 \begin{equation}
a_{\mu}=\frac{a_0}{4}, \qquad \alpha_{\mu}=\mu, \label{eq3}
\end{equation}
 the model describes a bistable system with fixed barrier height $V_{\mu}(0)=a_0/4$ but tunable minima positions:
 \begin{equation}
 y_{1,2}=\pm \frac{arcsinh(\mu)}{\mu}.
 \end{equation}
 \item For:
 \begin{equation}
 a_{\mu}=\frac{a_0\, \mu^2}{4\,(1+\mu^2)\alpha_{\mu}^2}, \qquad \alpha_{\mu}=arcsinh(\mu), \label{eq4}
 \end{equation}
 the model describes a bistable system with fixed minima positions i.e. $y_{1,2}=\pm 1$, but tunable barrier height, $V(0)= a_{\mu}$.
 \item For:
  \begin{equation}
 a_{\mu}=\frac{a_0\, \mu^2}{4\, arcsinh^2(\mu)}, \qquad \alpha_{\mu}=\frac{arcsinh(\mu)}{\sqrt{1+\mu^2}}, \label{eq5}
 \end{equation}
the model describes a bistable system with tunable minima positions i.e.:
 \begin{equation}
 y_{1,2}=\pm \sqrt{1 + \mu^2},
 \end{equation}
and tunable barrier height i.e.:
\begin{equation}
V_{\mu}(0)= \frac{a_0\, \mu^2}{4\, arcsinh^2(\mu)}.
\end{equation}
\end{enumerate}
Note that the quantity $a_0$ is real and positive, when $\mu\rightarrow 0$ the three members of the family of double-well potentials $V_{\mu}$ all reduce to:
\begin{equation}
V_0(y)=\frac{a_0}{4}(y^2 - 1)^2. \label{eq6}
\end{equation}
The latter is nothing else but the famous $\phi^4$ potential \cite{f4,dam1,f4a} used in several studies to investigate the dynamics of hydrogen-bonded molecular crystals \cite{a6,a10,a10a}, and structural phase transitions in ferroelectric crystals \cite{f4b,f4c}. In fig. \ref{fig1}, the effects of variations of the deformability parameter $\mu$ on the hyperbolic potential $V_{\mu}(x)$, for the three different sets of values of $a_{\mu}$ and $\alpha_{\mu}$ given by (\ref{eq3}), (\ref{eq4}) and (\ref{eq5}), are illustrated. We have set $a_0=0.6$.

The analytical solution of the nonlinear partial differential equation (\ref{eq8}) with the boundary conditions $y(z=x-vt)\rightarrow y_{1,2}$ and $\frac{\partial y}{\partial t}\rightarrow 0$ when $z\rightarrow \pm \infty$, is given by:
\begin{eqnarray}
y(x,t)&=&\frac{1}{\alpha_{\mu}}\,arc\tanh\biggl[\frac{\mu}{\sqrt{1+\mu^2}}\tanh\frac{x-vt}{2r_{\mu}(v)}\biggr], \label{eq9a} \\
r_{\mu}(v)&=&\frac{\mu}{2\alpha_{\mu}\sqrt{2(1+\mu^2)a_{\mu}}}\,r_0(v), \qquad r_0(v)=\frac{c_0}{\omega_0}\sqrt{1-\frac{v^2}{c_0^2}}. \label{eq9b}
\end{eqnarray}

\begin{figure}\centering
		\begin{minipage}[c]{0.34\textwidth}
\includegraphics[width=2.in, height= 1.8in]{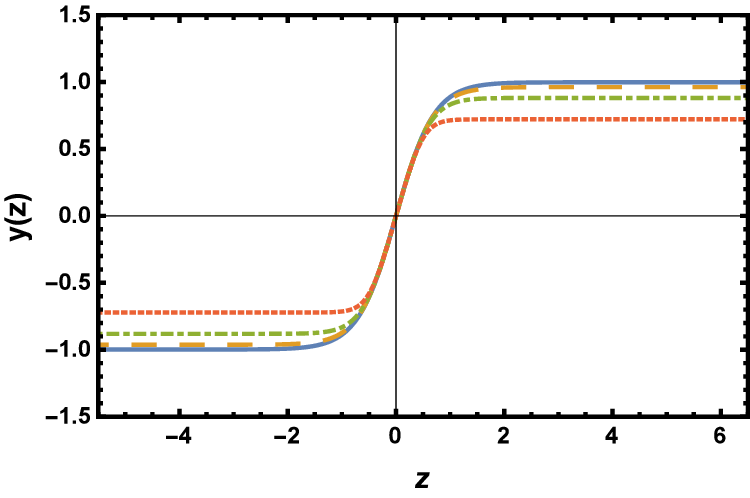}
		\end{minipage}%
		\begin{minipage}[c]{0.34\textwidth}
\includegraphics[width=2.in, height= 1.8in]{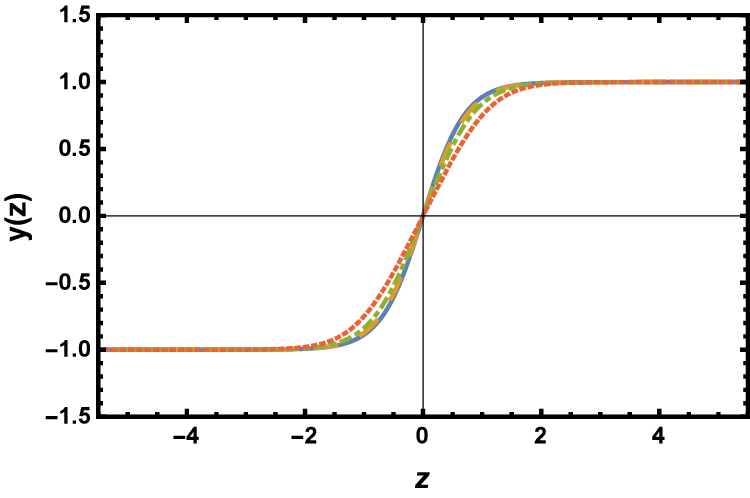}
		\end{minipage}%
		\begin{minipage}[c]{0.34\textwidth}
\includegraphics[width=2.in, height= 1.8in]{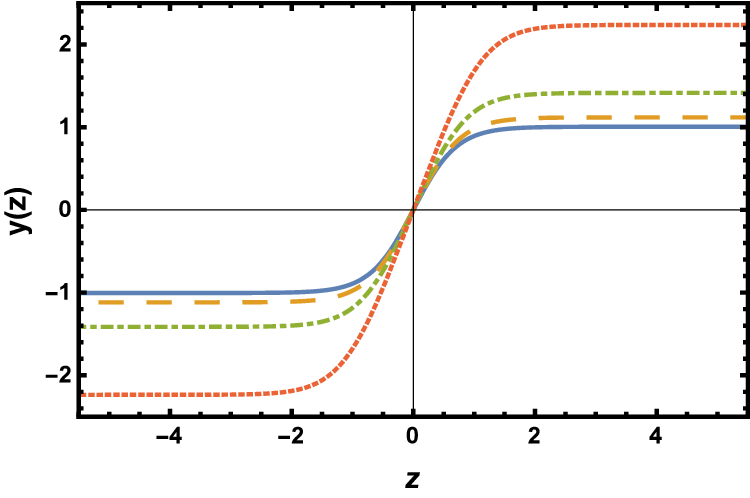}
		\end{minipage}
		\caption{Plots of $y(z)$ versus $z$ for $\mu=0.1$ (solid line),
   $\mu=0.5$ (dash line), $\mu=1.0$ (dash-dotted line) and $\mu=2.0$ (dot line). From left to right graphs: tunable minima but fixed barrier height, tunable barrier height but fixed minima, tunable minima and tunable barrier height.}
		\label{fig2}
	\end{figure}

\begin{figure}\centering
		\begin{minipage}[c]{0.34\textwidth}
\includegraphics[width=2.in, height= 1.8in]{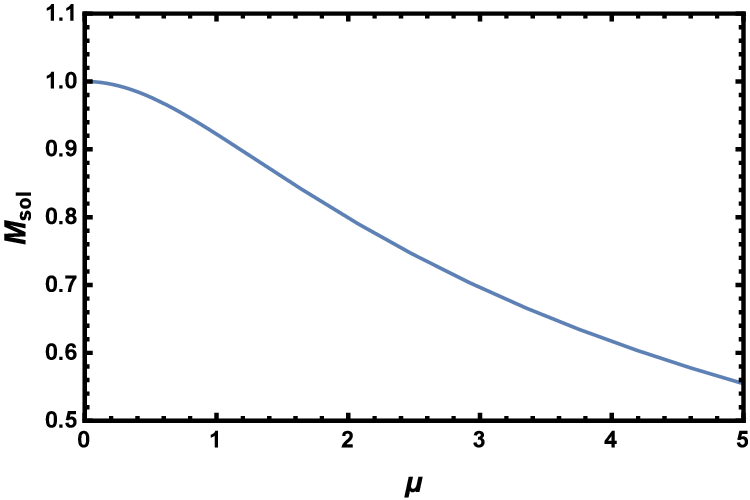}
		\end{minipage}%
		\begin{minipage}[c]{0.34\textwidth}
\includegraphics[width=2.in, height= 1.8in]{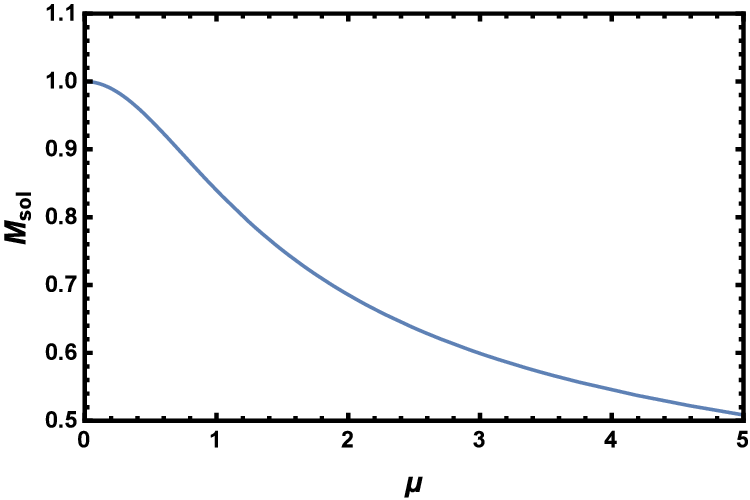}
		\end{minipage}%
		\begin{minipage}[c]{0.34\textwidth}
\includegraphics[width=2.in, height= 1.8in]{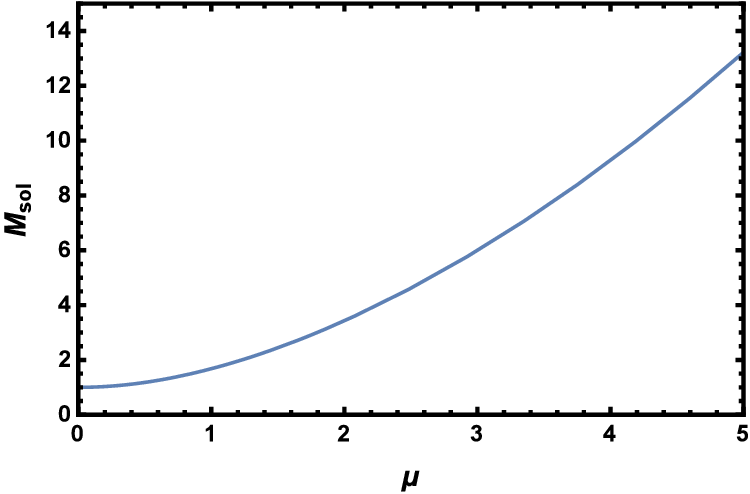}
		\end{minipage}
		\caption{Kink rest mass $M_{sol}(\mu)$ (in units of the $\phi^4$ kink rest mass $M_{sol}(0)$), plotted versus $\mu$. Left graph: double-well model with fixed barrier height but tunable minima, middle graph: double-well model with fixed minima but tunable barrier height, right graph: double-well model with tunable barrier height and tunable minima.}
		\label{fig3}
	\end{figure}

In fig. \ref{fig2}, the solution $y(z)$ is plotted versus $z$ for four different values of $\mu$. It is seen that $y(z)$ features a kink deformation created by molecular excitations in the infinitely long and continuous molecular crystal. The kink deformation propagates along the chain by translation at constant velocity $v$, with a width $r_{\mu}(v)$ that depends on the kink velocity. Note the direct proportionality of the kink width $r_{\mu}$ with the lattice spacing $d$, present in the expression of the sound speed $c_0$. Given that the kink solution (\ref{eq9a}) was obtained in the long-wavelength regime, its width $r_{\mu}$ should be far larger than the lattice spacing $d$. In fact for the kink to propagate by translation along the molecular chain, its width $r_{\mu}$ must be several times the intermolecular distance $d$. This is an absolute requirement for translational motion otherwise the kink could be pinned by the Peierls-Nabarro traps \cite{pe} erected on lattice sites, due to the discreteness of the molecular chain \cite{alph1,alph2}.
\\ The kink ampliude remains finite when $v=0$, suggesting a topolgical soliton. It is quite remarkable from figure \ref{fig2} that characteristic parameters of the kink defect (namely its width $r(\mu, v)$ and tail $y(0)$), are impacted distinctly for the three members of the family of hyperbolic double-well potential $V_{\mu}$ considered. Moreover, in addition to its topological nature, the kink solution eq. (\ref{eq9a}) behaves like a relativistic particle. By virtue of its particulate property the characteristic energy of the kink, obtained by integrating the continuous Hamiltonian (\ref{hcont}) with $y(z)$ given by eq. (\ref{eq9a}), reads:

\begin{equation}
E_{sol}(\mu, v)=\biggl(1 -\frac{v^2}{c_0^2}\biggr)^{-1/2}\,E_{sol}(\mu), \qquad
E_{sol}(\mu)= M_{sol}(\mu)\,c_0^2. \label{eners}
\end{equation}
$M_{sol}(\mu)$ is the kink rest mass and is defined:
\begin{equation}
M_{sol}(\mu)=\frac{m\,b^2}{2d\,\alpha^2_{\mu}\,r_{\mu}(0)}\,\int_{0}^{1}{\frac{(1-u^2)}{(1-b^2u^2)^2}\,du}, \qquad b=\frac{\mu}{\sqrt{1+\mu^2}}. \label{mass}
\end{equation}
In fig. \ref{fig3} the variation of the kink rest mass with $\mu$ is shown, for the three models of hyperbolic double-well potential.
We see that $M_{sol}(\mu)$ decreases as $\mu$ inscreases for the cases when either the barrier height is fixed or the minima positions are fixed, and increases with $\mu$ for the model where both the barrier height and the minima positions simultaneously change. Note a saturation of the kink rest mass to a finite lower threshold for the model with tunable minima but fixed barrier height, whereas  the kink rest mass seems to decrease monotonically toward zero for the model with tunable barrier height but fixed minima positions. This behaviour can be explained by the fact that a decrease of the barrier height creates less favorable conditions for the formation of kink defect in the system.

\section{\label{sec:level3}Kink-phonon scatterings}
\begin{figure}\centering
		\begin{minipage}[c]{0.34\textwidth}
\includegraphics[width=2.in, height= 1.8in]{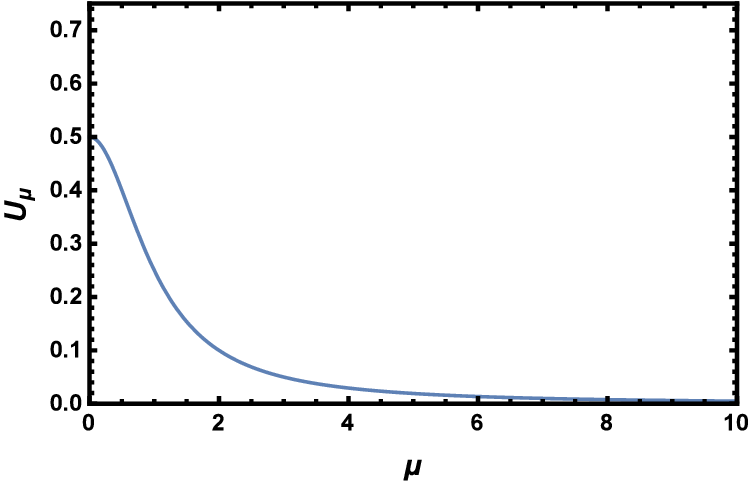}
		\end{minipage}%
		\begin{minipage}[c]{0.34\textwidth}
\includegraphics[width=2.in, height= 1.8in]{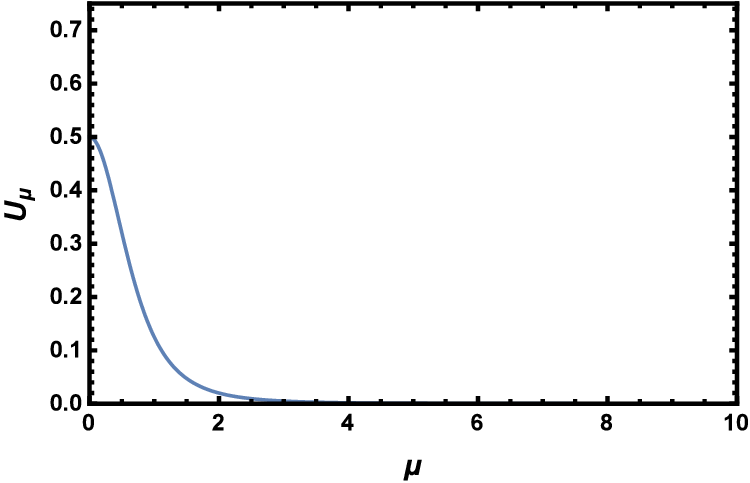}
		\end{minipage}%
		\begin{minipage}[c]{0.34\textwidth}
\includegraphics[width=2.in, height= 1.8in]{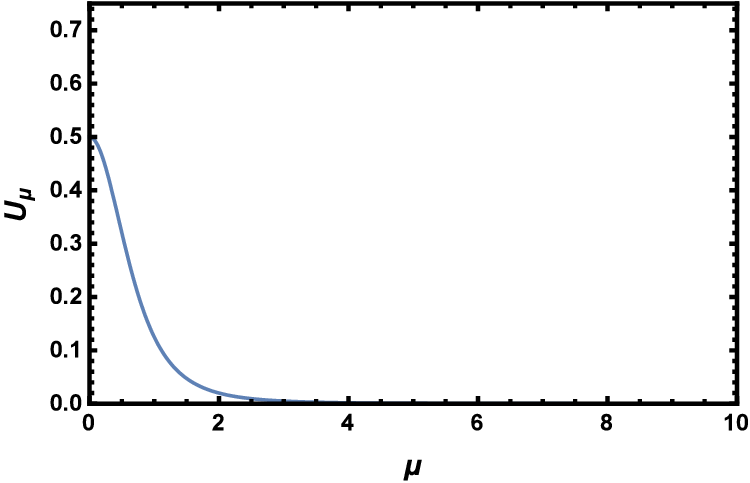}
		\end{minipage}
		\caption{Energy shift $U_{\mu}$ versus $\mu$. Left graph: double-well model with fixed barrier height but tunable minima, middle graph: double-well model with fixed minima but tunable barrier height, right graph: double-well model with tunable barrier height and tunable minima.}
		\label{fig4}
	\end{figure}

The kink response to small-amplitude excitations (i.e. phonons) is relevant for the analysis of kink stability in physical systems. Indeed in real physical contexts, a kink deformation will always coexist with small-ampliude oscillations trailing around and interacting with the kink. To analyze the interaction of kink with phonons for the nonlinear Klein-Gordon equation (\ref{eq8}), we rewrite $y(x,t)$ as:
\begin{equation}
 y(x, t)=y_0(x) + \psi(x, t), \label{ev1}
\end{equation}
where  $\psi(x,t)$ is the wavefunction of a perturbation propagating with the  kink. Equation (\ref{ev1}) suggests that we are interested in the stability of a static kink $y_0(x)$ given by (\ref{eq9a}) with $v=0$, coexisting with a perturbation of spatial amplitude $g(x)$ undergoing harmonic modulations in time i.e.:
\begin{equation}
\psi(x, t)=g(x)\,\exp(-i\Omega\,t), \label{solin}
\end{equation}
where $\Omega$ is the modulation frequency. Substituting the ansatz (\ref{ev1}) and (\ref{solin}) in eq. (\ref{eq8}) and expanding keeping only linear terms in $g(x)$, we are led to the following linear ordinary differential eqution in $g$:
\begin{equation}
 -c_0^2\,\frac{\partial g^2}{\partial x^2} + \omega_0^2\, U_{\mu}(x)\,g=\Omega^2\, g,  \label{ev2}
\end{equation}
where we defined $U_{\mu}(x)=\partial^2 V_{\mu}(y_0)/\partial y_0^2$. Setting $\nu=\Omega/\omega_0$ and introducing a new variable i.e. $\xi=x/r_0(0)$, eq. (\ref{ev2}) becomes:
\begin{equation}
 -\frac{\partial g^2}{\partial \xi^2} + U_{\mu}(\xi)\,g=\nu^2\, g.  \label{ev3}
\end{equation}
Equation (\ref{ev3}) is a linear eigenvalue problem with a scattering potential $U_{\mu}(\xi)$ given by:
\begin{eqnarray}
U_{\mu}(\xi)&=&U_{\mu}\bigg\lbrack 1 -\frac{3}{2}\,sech^2\,\frac{\xi}{2r_{\mu}} - \mu^2\,\biggl(sech^2\,\frac{\xi}{2r_{\mu}} - \frac{1}{2}\, sech^4\,\frac{\xi}{2r_{\mu}}\biggr) \bigg\rbrack, \label{pscata} \\
U_{\mu}&=&\frac{2a_{\mu}\alpha^2_{\mu}}{\mu^2(1+\mu^2)}, \qquad r_{\mu}= r_{\mu}(0)/r_0(0). \label{pscatb}
\end{eqnarray}
When $\mu\rightarrow 0$, $U_{\mu}(\xi)$ reduces to:
\begin{equation}
U_{0}(\xi)= \frac{1}{2}\,\biggl(1 -\frac{3}{2}\,sech^2\,\frac{\xi}{2r_0}\biggr), \qquad r_0=c_0/\omega_0, \label{pscatc} \\
\end{equation}
which is the $\phi^4$ scattering potential \cite{f4}. As a reminder the spectrum associated with the eigenvalue problem (\ref{ev3}) with the scattering potential (\ref{pscatc}), consists of discrete and continuous branches \cite{f4,dam1}. The discrete branch comprizes two boundstates, the lowest discrete mode exhibits the following signature:
\begin{equation}
\Omega_0=0, \qquad g_0(\xi)\propto sech^2\,\frac{\xi}{2r_0}.
\end{equation}
Since $\Omega_0=0$, the excitation of the lowest discrete mode entails no energy cost to the kink. However the scattering of this mode with the kink deformation will induce a uniform shift of the kink center of mass. For the latter reason this mode is often referred to as "kink translation mode" \cite{f4}. The second boundstate has:
\begin{equation}
 \Omega_1= \frac{\sqrt{3}}{2}\,\omega_0, \qquad g_1(\xi)\propto sech^2\,\frac{\xi}{2r_0} \sinh\,\frac{\xi}{2r_0}.
\end{equation}
In the general case when $\mu$ is nonzero we may rescale the dimensionless eigenfrequency $\nu$ by introducing:
\begin{equation}
\tilde{\nu}^2= \nu^2 - U_{\mu},
\end{equation}
and the eigenvalue equation (\ref{ev3}) turns to:
\begin{equation}
 -\frac{\partial g^2}{\partial \xi^2} + V_{\mu}(\xi)\,g=\tilde{\nu}^2\, g.  \label{ev4}
\end{equation}
In the latter eigenvalue equation, the scattering potential $V_{\mu}(\xi)$ is given by:
\begin{equation}
V_{\mu}(\xi)=U_{\mu}\bigg\lbrack - \frac{3}{2}\,sech^2\,\frac{\xi}{2r_{\mu}} - \mu^2\,\biggl(sech^2\,\frac{\xi}{2r_{\mu}} - \frac{1}{2}\, sech^4\,\frac{\xi}{2r_{\mu}}\biggr) \bigg\rbrack. \label{pscatd}
\end{equation}
This expession of $V_{\mu}(\xi)$ is similar to the potential proposed by Manning \cite{man}, in his study of energy levels of a symmetrical double-minima problem with applications to $NH_3$ and $ND_3$ Molecules. In the work of Manning \cite{man}, the energy levels of $V_{\mu}(\xi)$ were investigated using continued-fraction approximations combined with numerical simulations. Instructively several other approaches for solving the eigenvalue problem eq. (\ref{ev4}) have been proposed, most of them based on transformations to Heun's equation \cite{h1,h2,h3,h5,h6} or using series solutions in conjonction with algebraic Bethe ansatz \cite{h4}. \\
A more detailed analysis of eigenstates of eqs. (\ref{ev3}) and (\ref{ev4}) will be carried out in a separate work, however the lowest-energy states of the Manning potential (\ref{pscatd}) deserve useful comments. The first comment relates to its zero-energy mode $\tilde{\nu}=0$, which corresponds to the energy level $\nu= U_{\mu}$ of the scattering potential (\ref{pscata}). In fact $U_{\mu}(\xi)$ is just the Manning potential with a uniformly shifted spectrum. The discrete energy levels of the Manning potential $V_{\mu}(\xi)$ were investigated in ref. \cite{w1} by means of the Wronskian method, and results were compared with Manning's calculations \cite{man} on one hand, and with those obtained using the Riccati–Pad\'e method on the other hand. While a remarkable agreement was observed for results obtained using the Wronskian and Ricatti-Pad\'e methods, results from these two methods disagreed with those otained by Manning. The author \cite{w1} ascribed this discrepancy to inaccuracies in Manning’s calculations \cite{man}. Fig. \ref{fig4} shows the variation of the
energy shift $U_{\mu}$ with $\mu$, one clearly sees that the energy spectrum of the Manning potential $V_{\mu}(\xi)$ is generated by uniformly lowering the energy levels of the scattering potential $U_{\mu}(\xi)$ given in (\ref{pscata}).

\begin{figure}\centering
		\begin{minipage}[c]{0.34\textwidth}
\includegraphics[width=2.in, height= 1.8in]{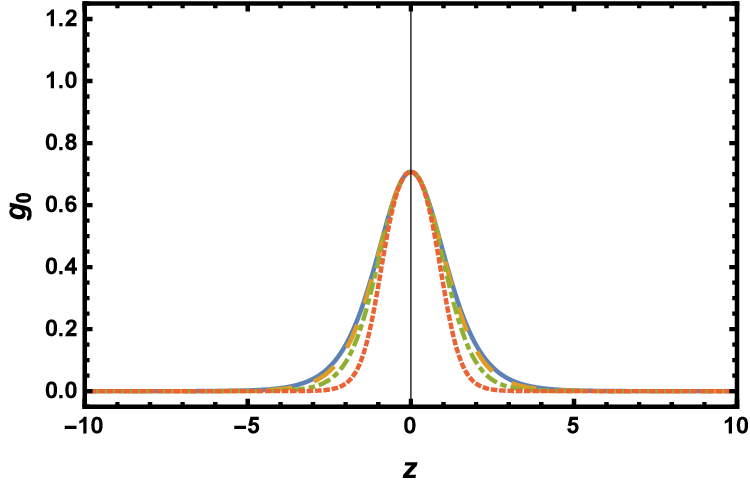}
		\end{minipage}%
		\begin{minipage}[c]{0.34\textwidth}
\includegraphics[width=2.in, height= 1.8in]{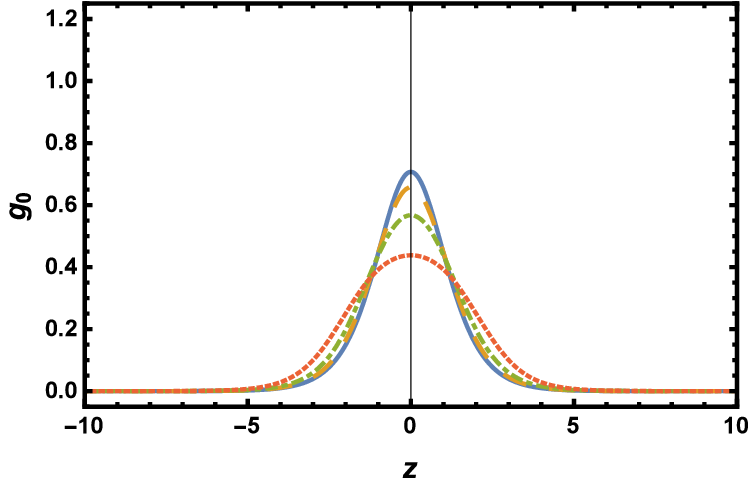}
		\end{minipage}%
		\begin{minipage}[c]{0.34\textwidth}
\includegraphics[width=2.in, height= 1.8in]{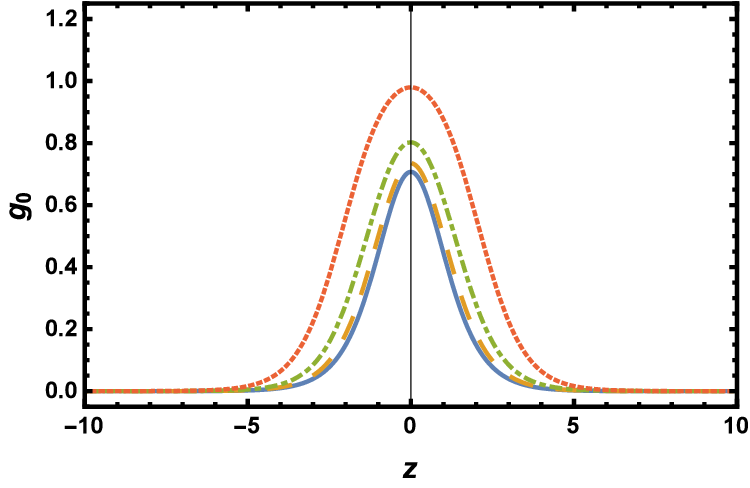}
		\end{minipage}
		\caption{Groundstate wavefunction $g_0$ given by (\ref{solmod}) versus $z$ for $\mu=0.001$ (solid line),
   $\mu=0.5$ (dash line), $\mu=1.0$ (dash-dotted line) and $\mu=2.0$ (dot line). From left to right graphs: variable minima but fixed barrier, variable barrier but fixed minima, variable minima and barrier.}
		\label{fig5}
	\end{figure}

The last but not the least relevant comment on the eingevalue problem eq. (\ref{ev3}), is about the eigenfunctions of some of its discrete modes that can be obtained analytically, contrasting with the Manning eigenvalue problem. Let us consider for proof its zero-energy mode $\Omega=0$. The associated eigenfuncton can straightforwardly be seen from eq. (\ref{ev3}) as being proportional to $d y/d \xi$, with a proportionality coefficient depending only on $\mu$. More explicitly the "Goldstone" translation mode of $U_{\mu}(\xi)$ exhibits the signature:
\begin{equation}
\Omega=0, \qquad g_0(\xi)\propto \sqrt{2a_{\mu}}\,\frac{(1+\mu^2)\,sech^2\,\frac{\xi}{2r_{\mu}}}{1+\mu^2 sech^2\,\frac{\xi}{2r_{\mu}}}. \label{solmod}
\end{equation}
In fig. \ref{fig5} the wavefunction $g_0$ is seen to be pulse-shaped for the three models, irrespective of values of $\mu$.
$U_{\mu}(\xi)$ thus emerges to be an exactly integrable potential, unlike the Manning potential for which
only approximate solutions can be obtained \cite{man,w1}.

\section{\label{sec:level4}Conclusion and perspectives}
Kinks are structural defects that form in low-dimensional molecular systems as a result of the competition between dispersion and nonlinearity. In these systems the dispersion is brought about by intermolecular interactions, while nonlinearity is due to a one-body force created by a substrate potential. Kink deformations have been shown to ensure proton transfers along hydrogen-bonded molecular chains \cite{bo1,bo11}, they are the most common forms of topological deformations \cite{bo2,bo3} in the structures of molecular crystals characterized by an inversion symmetry. Indeed molecular excitations in the latter systems involve a bistable configurational energy wich are described by a double-well potential. Configurational energies with double-well landscapes are equally relevant in the folding dynamics of protein chains during their conformational transitions, notably in the case of staphylococcal protein $\alpha$ it has been observed \cite{alph1,alph2} that such configurational energy favors the formation of kink defects as the result of the winding-unwinding dynamics of protein's $\alpha$ helix. In these biochemical and  biophysical systems where the configurational energy is characterized by a bistable symmetry, the shape deformability of the double-well potential stems from isotope effects such as ionic substitutions that can change the equilibria of individual molecules, in order to adjust to their new environments. In mathematical physics a kink is a solution to the nonlinear Klein-Gordon equation, it possesses particulate features and by vertue of its topological nature it can be endowed with a relativistic energy, and hence a relativistic mass.\\
In this work we investigated the interactions of kinks with small-amplitude excitations, for a nonlinear Klein-Gordon model characterized by a bistable on-site substrate represented by a family of hyperbolic double-well potentials. The double-well shape of this family of hyperbolic substrates was assumed tunable through one single control parameter: for one member of the family, the two degenerate minima can be shifted continuously without affecting the height of the potential barrier. In a second member the two potential minima are always fixed while the height of the potential barrier can be continuously tuned. For the third member, both the two minima and barrier height can be tuned simultaneously. Since the nonlinear Klein-Gordon Hamiltonian for the molecular chain was not tractable analytically due to its discrete nature, we carried out a continuum-limit approximation and derived a continuous nonlinear Klein-Gordon equation, that led to an analytical expression for the kink-soliton solution for the family of hyperbolic double-well substrate potentials. From the continuous nonlinear Klein-Gordon Hamiltonian, we derived the kink relativistic energy and subsequently the kink relativistic mass. The kink scattering with small-amplitude excitations or phonons, was examined by constructing the corresponding eigenvalue problem. The scattering potential obtained for this linear eigenvalue problem was shown to reduce, after rescaling, to the potential proposed by Manning \cite{man} in the study of energy levels of $NH_3$ and $ND_3$. The exact analytical expression of the groundstate wavefunction for the kink-phonon scattering eigenvalue problem was obtained, suggesting that the corresponding scattering potential is exactly integrable. It is remarkable that the Manning potential so far has not been solved exactly, most of its existing solutions were obtained using approximate methods combined with numerical simulations \cite{w1}. \\
As mentioned earlier, the discrete nature of the nonlinear Klein-Gordon Hamiltonian (\ref{eq2}) makes it difficult to solve analytically the corresponding nonlinear Klein-Gordon equation. Although the continuum-limit approximation enabled us to obtain an analytical kink solution, this solution remains approximate suggesting that the eigenvalue equation (\ref{ev2}) too would provide only an approximate picture of the kink-phonon scattering problem for the discrete molecular chain. Still the continuum-limit consideration is not just an approximation in several systems, instead it reflects the actual physical context. This is notably the case in molecular chains for which the lattice spacing is too small compared with the typical wavelengths of molecular excitations. Instructively, the asumption that the molecular chain behaves like a continuous medium implies that the kink propagates along the chain without feeling the Peierls-Nabarro stress created by lattice disretenss. In fact the kink will uniformly translate along the molecular chain, suggesting that the zero-energy mode resulting from its scattering with phonons (i.e. Goldstone mode) induces a uniform shift of the kink center-of-mass position by a distance proportional to the amplitude of the mode wavefunction. This is to say that the Goldstone mode is consequent upon a continuous-symmetry breaking in the translational motion of the kink soliton, a phenomenon that is no expected in the discrete regime where the kink motion occurs via jumps across the Peierls-Nabarro barriers erected on lattice sites \cite{pe,alph1,alph2}.\\
In this study we considered only the zero-energy mode of the potential generated by the kink-phonon scattering. A detailed analysis of the discrete and continuous branches of this spectrum, is useful in order to gain insight onto the stability of kinks \cite{bo2,bo3} in systems concerned by the proposed model.

\begin{acknowledgments}
The author is endebted to the Alexander von Humboldt foundation, for supporting his visit at Max-Planck Institute for the Physics of Complex Systems (MIPKS) in Dresden, Germany, where this study was initiated. I wish to address special thanks to Holger Kantz for making everything possible for a smooth and comfortable stay in the "Nonlinear Dynamics and Time Series Analysis" reasearch unit.
\end{acknowledgments}

\section*{Author declarations}
\subsection*{Conflict of Interest}
The author has no conflicts to disclose.
\subsection*{Author Contributions}
A. M. Dikand\'e: Conceptualization (lead); Formal analysis (lead); Investigation (lead); Methodology (lead); Writing original draft (lead).
\subsection*{Data availability}
Data sharing is not applicable to this article, as no new data were generated in the study.

\end{document}